# Gaussian Equation to Describe the Percent of Shadow Length in Satellite Image

Emad Ali Al-Helaly[1],    Israa J. Muhsin[2]

[1]Department of Civil, College of Engineering, Kufa University, [2]Department of Remote Sensing and GIS, College of Science, Baghdad University
E-mail: emaadalhilaly@gmail.com ; imada.alhilali@uokufa.edu.iq

**Abstract:**

The shadow is a separated feature in the satellite images specially high resolution images. It may be looked as a problem when it causes a loss in the ground ray response to the shaded area, and may be, from other view, considered as an indication to the height of the bodies or buildings, when it hidden due to top view of the satellite image usually.

The location of the object on the earth affects its shadow length, direction and darkness. The ostensible sun path in the sky also affects the shadow length and direction, it is depending on the change of the rotation axis of planet about its center during the year. And it is clear that the shadow is changing with the sun transmission in the day times.

The sun location needs complex equations, because there are several angles describing its location with taking into consideration the location of the shaded body on the earth spatially the latitude, and the shadow direction and length depend on the sun location so it need more equations.

Some engineering purposes, and some engineers and designer can not process these complex astronomy equations every time they need to calculate the shadow effect on the urban region or garden for examples, or other architectural purposes, it is unpractical process to estimate the shadow direction and length for civil from these equation.

So, we derived in this study a high accurate empirical Gaussian equation to find shadow length percent in the limits of the study area depends on single parameter which is the number of day during the year, and calculating the percent of shadow, so we can calculate the shadow length percent accurately in any day number by the suggested equation after knowing the real building height.

## 1. Introduction: light and shadow

Light is the electromagnetic spectrum rays in the visible range which coming usually from the sun. The electromagnetic rays propagate in straight lines from the light source.

The opaque objects obscure light to reach area in the opposite side if the light source, so the shadow area still dark relatively and called the shadow. So, the shadow is the darkness that caused by an object when it blocks the light from getting a corresponding surface [R.1]. We can say also, the shadow occurs in the image when there is an object that obscures the light coming from the light source completely or partially [R. 2].

## 2. Previous works

The phenomena of the shadow was appeared clearly after 2000 with the increasing of resolution satellite sensors (IKONOS, Quick-Bird, WorldView, and other), so many studies were written about shadow to remedy or use it as a function to know the properties of shaded body.

C. Lin and R. Nevatia, 1998, adopted complex algorithms to estimate the three-dimensional model from the two-dimensional image [R. 4].

K. Karantzalos and N. Paragios, 2008, were adopted a metadata or ground information to find the hidden dimension, i. e. height of the objects in a simple way [R. 5].

Won Seok, et. al, 2007 suggested geometric method for estimate the building (height) from one satellite image separately, he depended some astronomic parameters in the remote sensing imaging system as solar angles, image scale, direction of the shadow, angle of satellite imaging [R. 3].

V. Arévalo et. al. determined the shadow in images of high resolution satellite Quick-Bird of resolution (0.6 meters), but said it is a way suitable for images satellites IKONOS and WorldView. In all the angle of the Sun was taken fixed, and V. Arévalo et. al supposed that the calculating of sun angle at over particular day along the year is very complex [R. 6].

Using SPOT satellite images V. K. Shettigara and G. M. Sumerling, 1998 calculated tree and buildings heights, and succeeded in calculating the height of the buildings accurately, but they were not high accuracy in the calculation of tree height results due to informal form in the trees [R. 7].

## 3. The shadow geometry

The shadows are two dimensional forms that result from three dimensional dark bodies, and depend on the geometrical properties of these bodies, and at the angle of incident rays of the sun.

**Angle of the sun and its synchronization**

Sun is the main source of all electromagnetic waves directly or indirectly, the sun appears moving in the sky along the day, and the shadow is changing according to the sun direction.

The angle of the sun is varying all the year and day hours because of the difference in the angle of the fall of radiation on the ground.

In addition, it is seemed moving in complex orbit on the sky from the observer on the Earth.

The incident angle of the sun light on the Earth's surface varies by distance from the equator because of the spherical shape of the planet. The sun is at 21/22 of March perpendicular to the equator so that there are no shadows to the vertical structures there, but it will be inclined in areas northern or southern equator due the spherical shape of the planet. Here we will discuss the relation between the sun angle and the scene on the earth, which affect the shadow length and direction.

**The sun**

The sun's diameter is about 1,390,000 kilometers [R. 8] and the average distance to the Earth's surface is about 149.5 million kilometers. It generate various wavelengths of electromagnetic radiation from long waves of radios and short wavelength of cosmic rays.

Electromagnetism which coming from sun consists of approximately 46 % visible radiation and 46 % thermal radiation [R. 9]

The Earth is rounding at a biosphere orbit around the sun, though, the amount of rays intensity falling on the ground varies throughout the year. The earliest distance to the Earth is 147 million km at the beginning of January, and further distance is 152 million km at the beginning of July [R. 10].

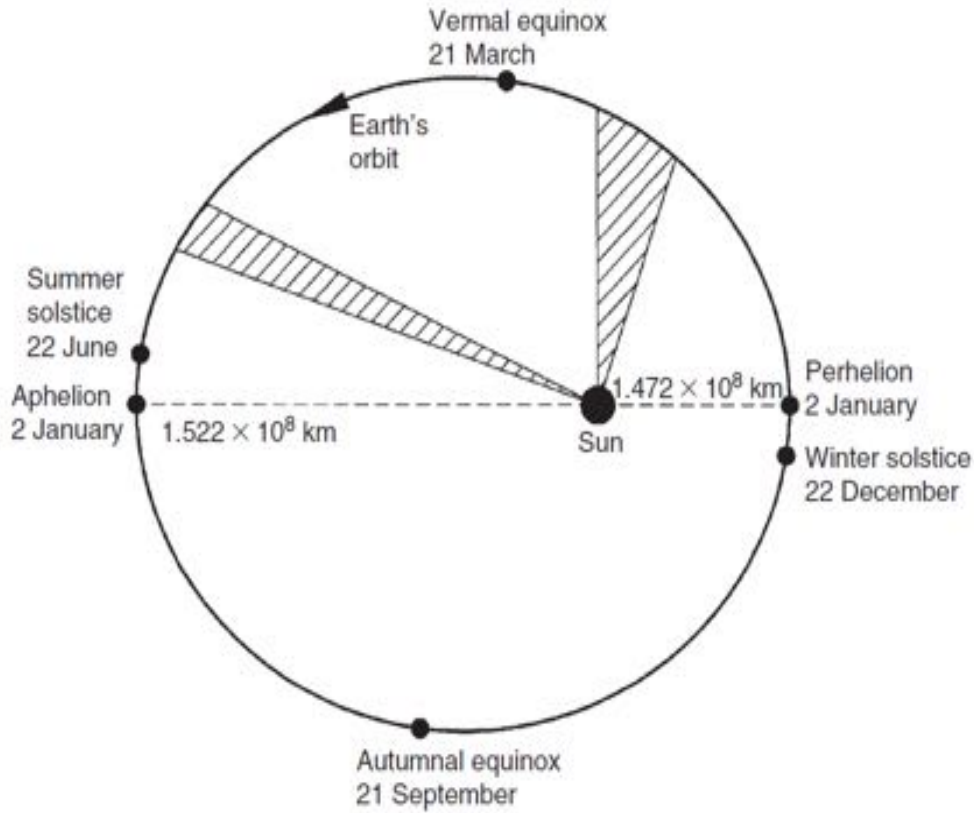

Figure (1): Earth's orbit around the sun [R. 11].

Accurately the distance [R11, p. 3] between the sun and earth $(r_o)$= 1.496 *$10^8$ km, or in more accurate 149597890±500 km [R. 11, p. 3], and (r) is the sun-earth distance for every day in the year.

The day angle ($\Gamma$ : rad) is calculated from the equation below:

$\Gamma = 2\pi (d_n-1)/365$ ………………………………………..(1)

Where:
$d_n$ is the day number, it is 1 to the 1- January, and 32 to 1-February.

The Earth's axis of rotation across the year is varying to be about ±23.5 degrees [R. 11] during the year, which affects the sun incident angle and it's difference on the surface of the earth, thus affects the direction of the shadow and its extension.

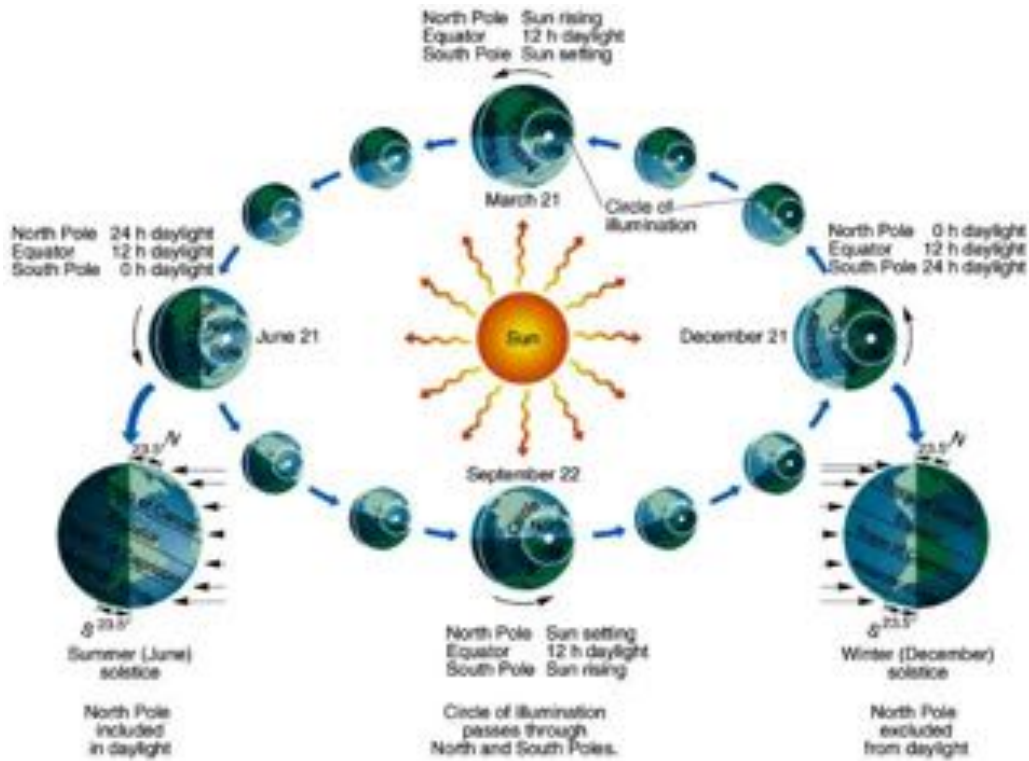

Figure (2): The changing in the Earth's axis rotation around itself during the year [R. 10]

The angle, position and elevation of the sun can be organized by astronomical equations [R. 11] in the following manner

**Sun location**

**The angle of solar deviation (δ):** it is the angle between the equatorial plane (the plane of Earth revolution about the sun) and the line between the center of the Earth and the center of the sun. It is also termed declination angle. The value of this angle ranges between +23.45 and -23.45, and the value of this angle varies day to day. At 20/21 March and 22/23 September it equals zero. It is calculated from the relationship:

$$\delta = (0.006918 - 0.399912*\cos(\Gamma) + 0.070257*\sin(\Gamma) - 0.006758*\cos(2\Gamma)$$

$$+ 0.000907*\sin(2\Gamma) - 0.002697*\cos(3\Gamma) + 0.00148*\sin(3\Gamma)) *(180/\pi)$$

.....................................(2)

This equation is came from a Fourier formula, was developed by J. W. Spencer [R. 12], and it is the most accurate equation. There are other simple equation like [R. 13]:

$$\delta = \pi*18023.45* \sin(2\pi*365284 + d_n) \quad \ldots\ldots\ldots\ldots\ldots\ldots (3)$$

Or this equation:

$$\delta = \sin^{-1} * \sin(23.450) * \sin(2\pi * d_n - 81365) \ldots\ldots\ldots\ldots..(4)$$

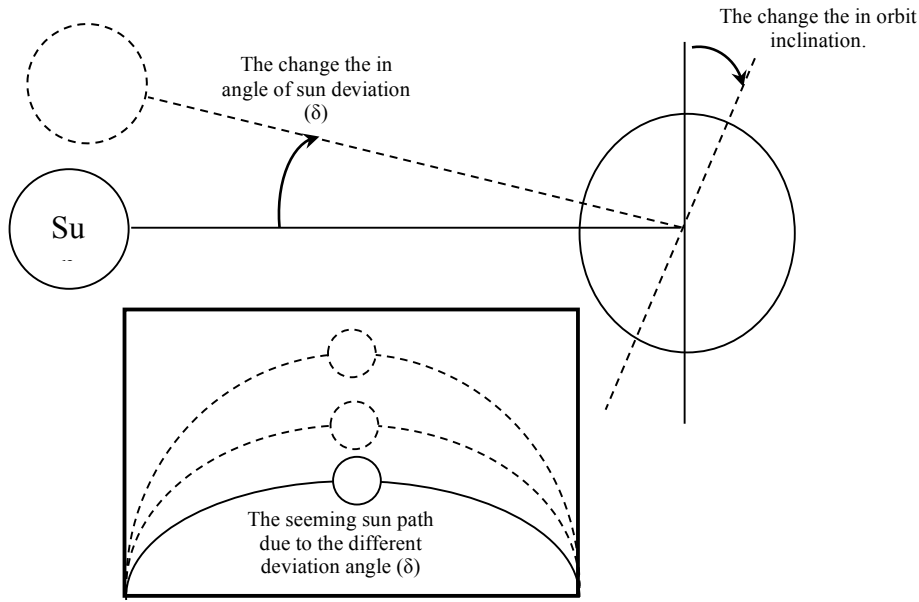

Figure (5) A sketch to the sun seeming deviation angle.

**The geographic latitude angle (ϕ):** this is the earth point coordinate, it is positive to the north of the Earth, and zero at the equator line.

The shadow of all objects is lengthened words toward the poles of the planet, even if they are in the same latitude.

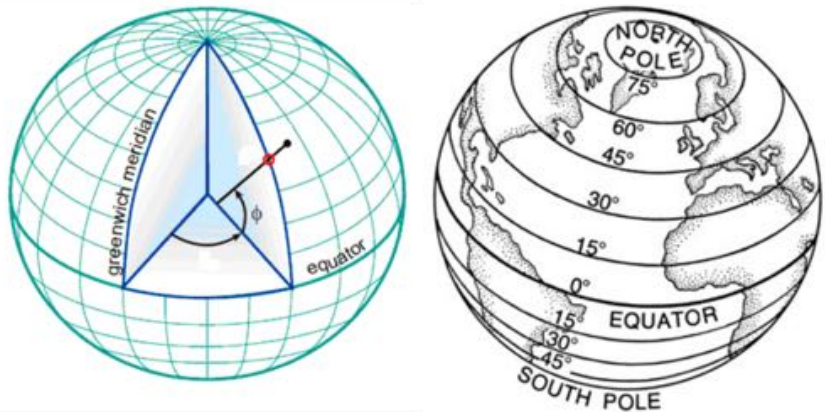

Figure (6) The geographic latitude angle (ϕ).

**The hour angle (ω):** the hour angle measured from the zenith line to the horizon. It is zero at noon, and with positive value along the morning. At the sine shine, it equals 90°, and it is distributed equally with the hours of the day half. The same graduating after noon but in negative values.

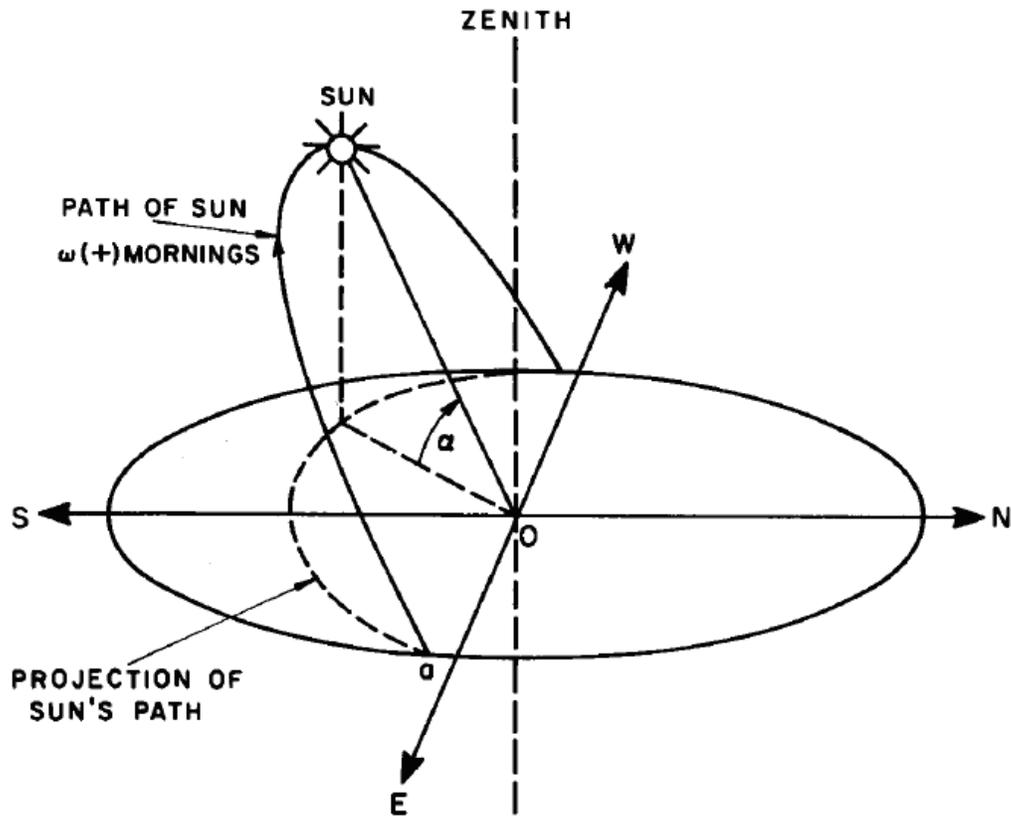

Figure (7) The hour angle (ω).

**The angle of the solar rise (α):** is the angular height of the sun in the sky measured from the horizon (the horizon of the observation with the Earth is the circle obtained from the intersection of the sky of the observer with the Earth). This angle is equal to zero at sunrise and sunset, and 90 when the sun is directly above the observatory. It also may be named solar elevation angle.

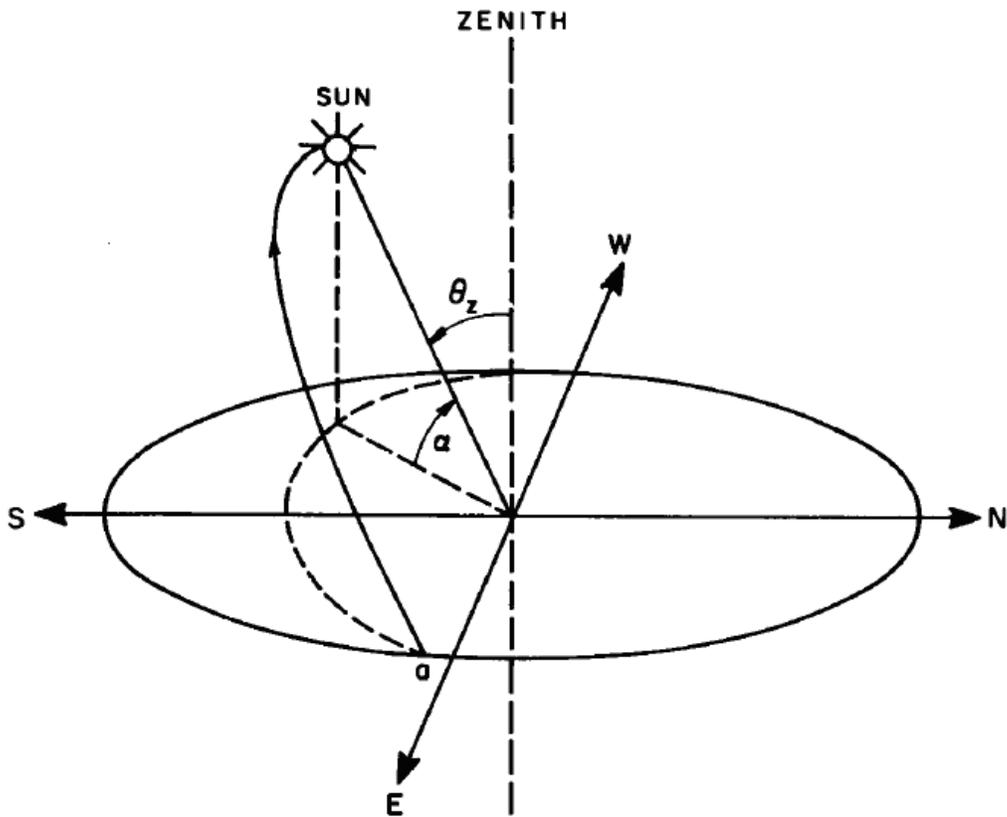

Figure (8) The angle of the solar rise (α).

**The azimuth angle ($\theta_z$):** is the complementary angle of the angle of elevation (α), and measured from the roots, which is after the center of the sun degrees from the point of the head (Zenith), and roughly the angle of the head's head at sunrise or sunset is 90 degrees. It is also named (zenith angle), and can be calculated from the equation below [R. 11, p. 15]:

$$\alpha = 90 - \theta_z \quad \text{................(5)}$$

The elevation angle of the relationship is calculated by [R. 11, p.15]:

$$\cos(\theta z) = \sin(\alpha) = \cos(\phi) * \cos(\delta) * \cos(\omega) + \sin(\phi) * \sin(\delta)$$

$$\text{................(6)}$$

Now, we can calculate every sun coordinate, or angles depending, and shadow on the number of days for everybody when its orientation and dimensions are known.

## 4. The selected building

The selected building is a (Qasr Al-Safer hotel and restaurant) in Najaf, City. The orientation of it is ($32^o$ 00' 02.49" N) and ($44^o$ 21' 32.88" E). or (32.000691 N 44.359133 E).

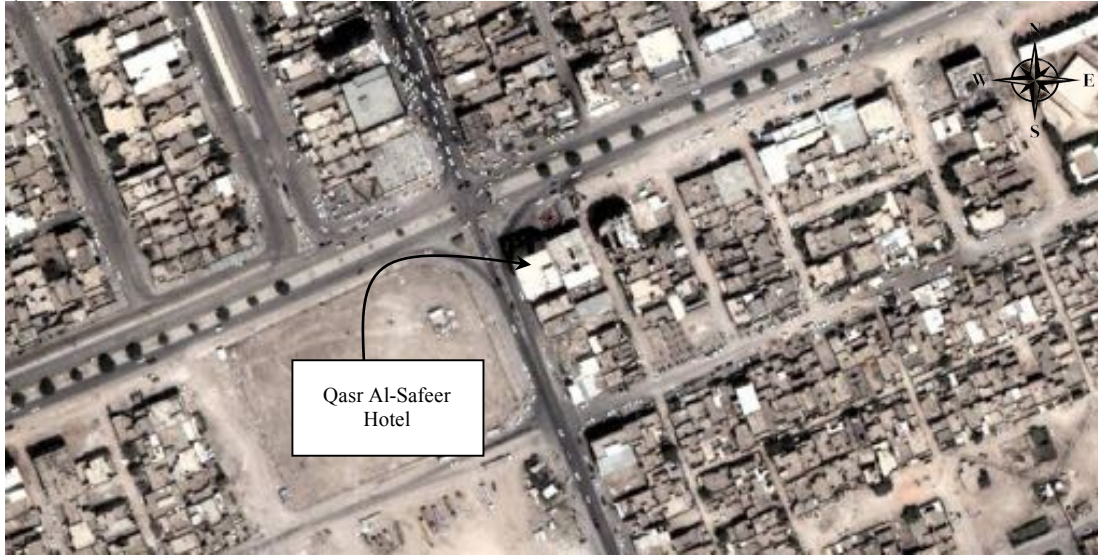

Figure (3) The orientation of Qasr Al-Safeer hotel and restaurant.

The image above is a band composed Quick-Bird imagery by WGS 84 map projections [R. 14]. And the seen above is imaged at 09:43 a.m. according to the key information file which is attached with the imagery CD.

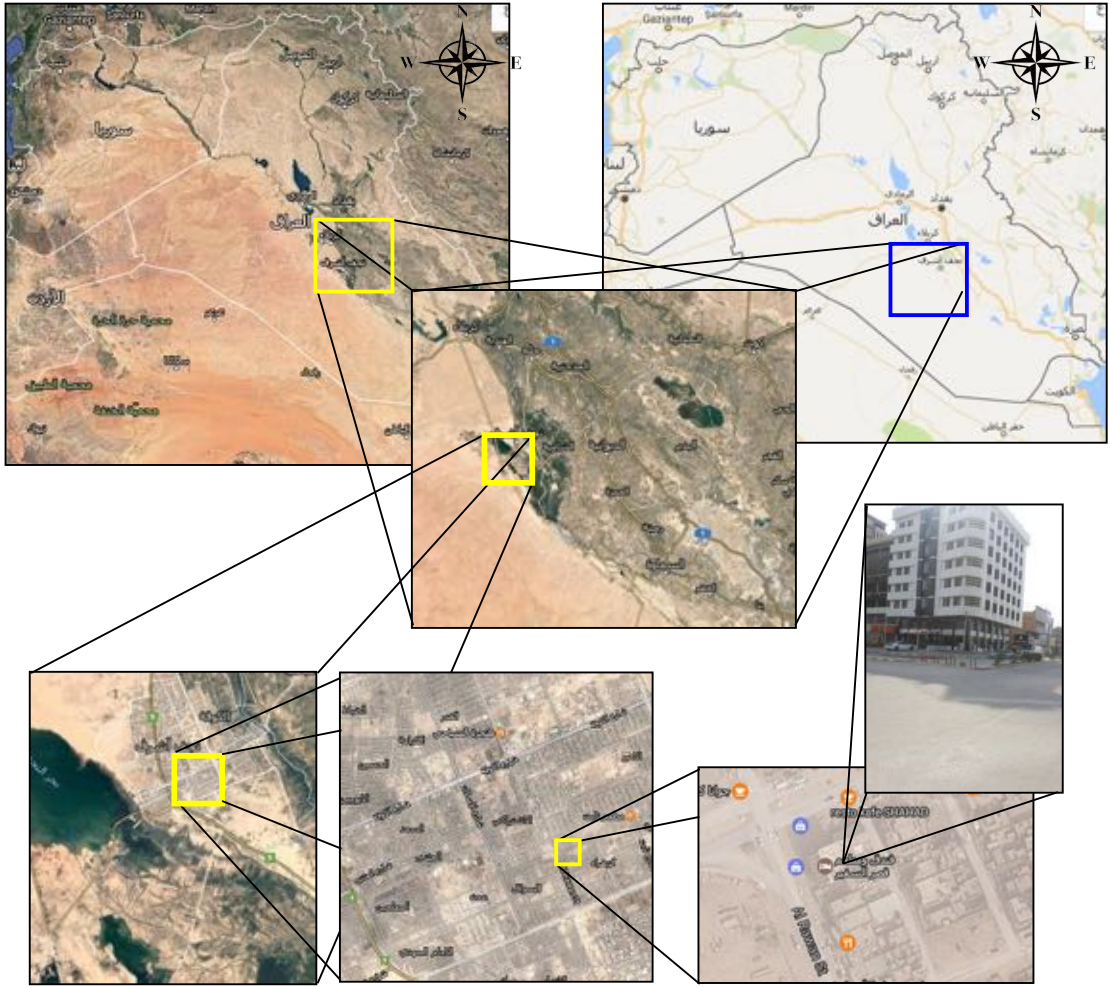

Figure (4) The location of the selected building.

The building is 25.30 m height over the adjacent street level, on the crossing of two streets (Al-Rawan and Al- Ameer), and there are no high buildings around it except at one side. These properties make easy to measure the length of the shadow from two sides. By the electronic and magnetic composes, we found the direction of building walls to measure the direction of shadow according to it, by this we need no repeat the field measurements of shadow according to the eastern direction.

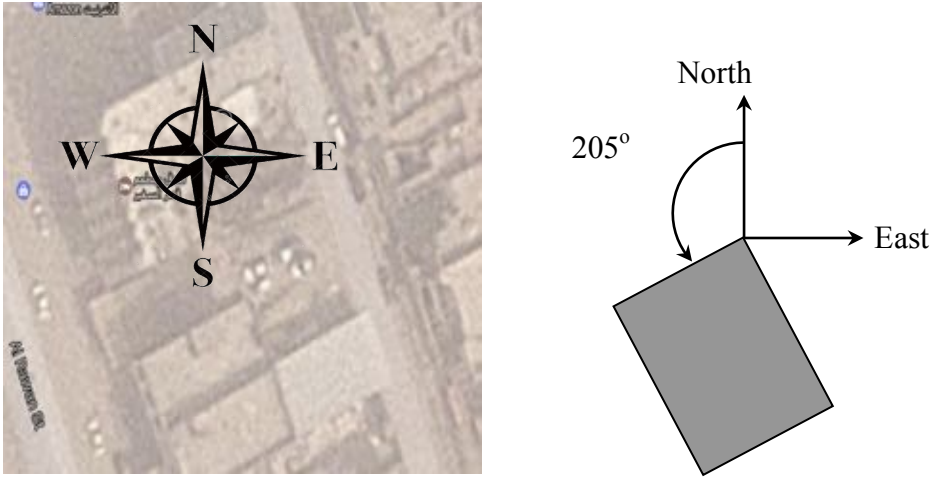

Figure (5) The direction of the walls of the selected building.

## 5. The sun orientations during the year

The sun orientation in the sky is calculated from astronomy equations during the years as presented.

For the Quick-Bird imaging time at 29/8/2006:
The day number ($d_n$) is 241.

The deviation angle ($\delta$) from equation (2)

$\Gamma= 2\pi(d_n-1)/365 = 4.131409517$ rad

$\delta=(0.006918-0.399912*\cos(4.13141)+0.070257*\sin(4.13141)-$

$0.006758*\cos(2*4.13141) +0.000907*\sin(2*4.13141)-0.002697$

$* \cos(3*4.13141) +0.00148*\sin (3*4.13141) *(180/\pi)$

$\delta= 0.168285459$ rad

$= 9.642046551$ degrees

The latitude angle ($\phi$) is + 32.000691 degrees. From the latitude circle

The hour (solar) angle ($\omega$)= 12:00- 09:43= 02:17 hours = 34.25 degrees

$= 0.711$ rad

The zenith angle ($\theta_z$) from eq. (6)

$\cos(\theta_z)=\sin(\alpha)= \cos(\phi) * \cos(\delta) * \cos(\omega) + \sin(\phi) * \sin(\delta)$

$$= \cos(32) * \cos(9.642) * \cos(34.25) + \sin(32) * \sin(9.642)$$
$$= 0.848*0.98587*0.82658975+0.52992*0.16749$$
$$= 0.7798$$

$\theta_z = 38.7577$ degrees

The angle of solar rise ($\alpha$)= 51.24 degrees from eq. (5)

From this angle we can calculate the shadow length:

tan ($\alpha$)= building height /shadow width (measured from field or image at that date: 29/August)

$$= H/ 19.95 \text{ m}$$

H= 19.95*tan(51.24227)= 24.85 m

But from the field the height of building is 25.30 m

So, if we calculate the shadow geometrically from the astronomic equation with known height.

tan ($\alpha$)= 25.3/shadow length

shadow length= 25.3/tan ($\alpha$)= 20.31 m

It is easy to program these calculations by any programming language. Then we make a table to all shadow length.

Table (1) Shadow length and direction by the astronomy equations (rounded values to two digits).

| date | day number | day angle | delineation angle | latitude angle | hour angle | azimuth angle | solar rise angle | shadow length (m) |
|---|---|---|---|---|---|---|---|---|
| 29-Aug | 241 | 4.13 | 0.17 | 32 | 34.25 | 0.68 | 51.25 | 20.31 |
| 13-Sep | 256 | 4.39 | 0.07 | 32 | 34.25 | 0.74 | 47.49 | 23.19 |
| 28-Sep | 271 | 4.65 | -0.03 | 32 | 34.25 | 0.82 | 43.23 | 26.91 |
| 13-Oct | 286 | 4.91 | -0.13 | 32 | 34.25 | 0.89 | 38.77 | 31.50 |
| 28-Oct | 301 | 5.16 | -0.22 | 32 | 34.25 | 0.97 | 34.43 | 36.90 |
| 12-Nov | 316 | 5.42 | -0.31 | 32 | 34.25 | 1.04 | 30.61 | 42.76 |
| 27-Nov | 331 | 5.68 | -0.37 | 32 | 34.25 | 1.09 | 27.68 | 48.23 |
| 12-Dec | 346 | 5.94 | -0.40 | 32 | 34.25 | 1.12 | 25.98 | 51.93 |
| 27-Dec | 361 | 6.20 | -0.41 | 32 | 34.25 | 1.12 | 25.70 | 52.57 |

| date | day number | day angle | delineation angle | latitude angle | hour angle | azimuth angle | solar rise angle | shadow length (m) |
|---|---|---|---|---|---|---|---|---|
| 11-Jan | 11 | 0.17 | -0.38 | 32 | 34.25 | 1.10 | 26.90 | 49.88 |
| 26-Jan | 26 | 0.43 | -0.33 | 32 | 34.25 | 1.06 | 29.42 | 44.86 |
| 10-Feb | 41 | 0.69 | -0.26 | 32 | 34.25 | 0.99 | 33.00 | 38.96 |
| 25-Feb | 56 | 0.95 | -0.16 | 32 | 34.25 | 0.92 | 37.25 | 33.28 |
| 12-Mar | 71 | 1.20 | -0.06 | 32 | 34.25 | 0.84 | 41.77 | 28.33 |
| 27-Mar | 86 | 1.46 | 0.04 | 32 | 34.25 | 0.76 | 46.19 | 24.27 |
| 11-Apr | 101 | 1.72 | 0.14 | 32 | 34.25 | 0.69 | 50.18 | 21.09 |
| 26-Apr | 116 | 1.98 | 0.23 | 32 | 34.25 | 0.64 | 53.49 | 18.72 |
| 11-May | 131 | 2.24 | 0.31 | 32 | 34.25 | 0.59 | 55.97 | 17.08 |
| 26-May | 146 | 2.50 | 0.37 | 32 | 34.25 | 0.57 | 57.60 | 16.06 |
| 10-Jun | 161 | 2.75 | 0.40 | 32 | 34.25 | 0.55 | 58.45 | 15.54 |
| 25-Jun | 176 | 3.01 | 0.41 | 32 | 34.25 | 0.55 | 58.64 | 15.42 |
| 10-Jul | 191 | 3.27 | 0.39 | 32 | 34.25 | 0.56 | 58.19 | 15.69 |
| 25-Jul | 206 | 3.53 | 0.35 | 32 | 34.25 | 0.57 | 57.06 | 16.39 |
| 9-Aug | 221 | 3.79 | 0.28 | 32 | 34.25 | 0.61 | 55.12 | 17.64 |
| 24-Aug | 236 | 4.05 | 0.20 | 32 | 34.25 | 0.66 | 52.35 | 19.52 |

## 5. The fields measurements

A scale tape and simple instruments achieved the field measurements.

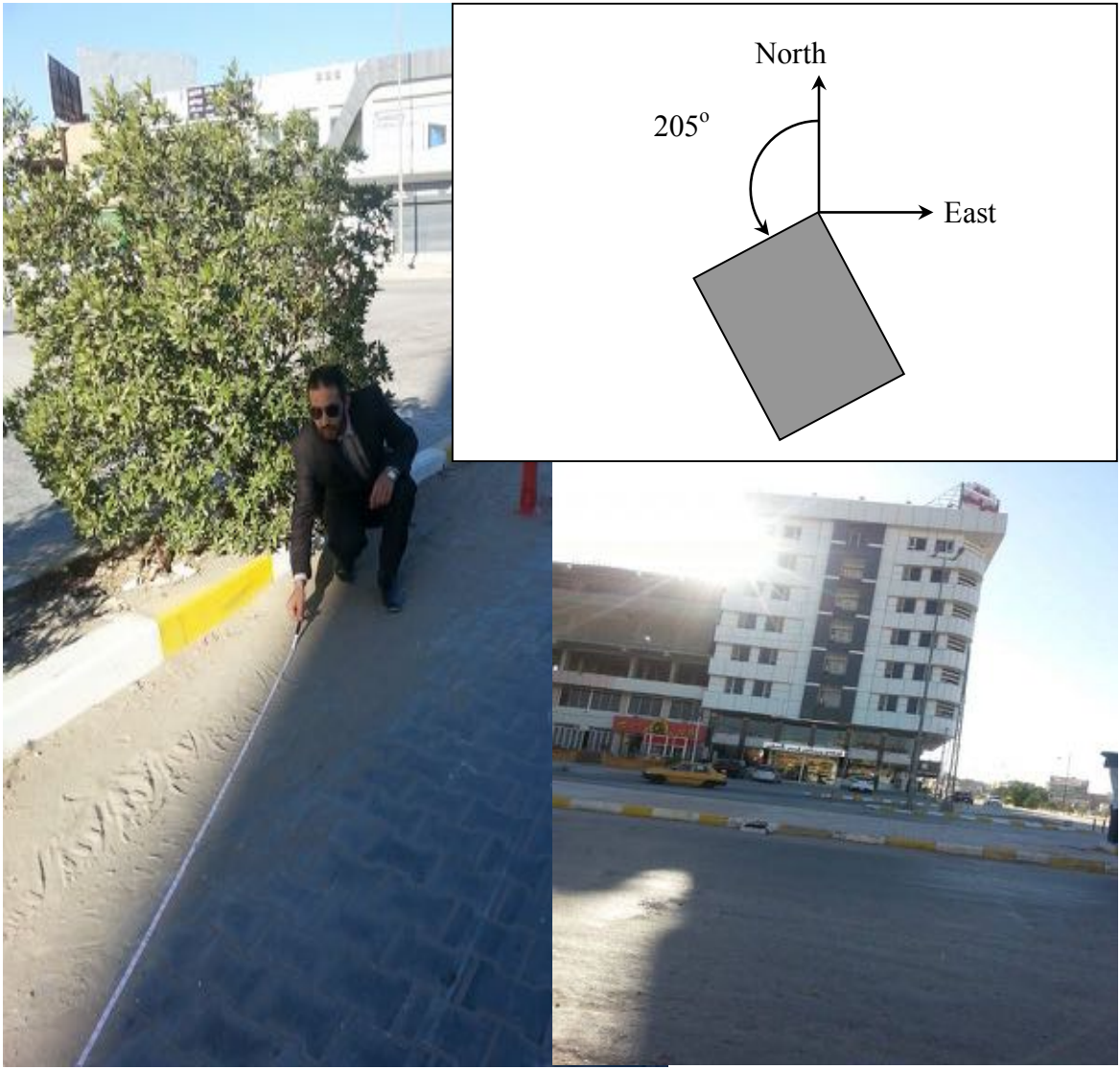

Figure (6) Field measurements of shadow length and direction.

The results are illustrated in the table below:

Table (2) Illustrate the field measurements for the length and direction of building shadow.

| Date | Shadow angle from the North (degrees) | shadow length (m) |
|---|---|---|
| 29-Aug | 51 | 20.20 |
| 13-Sep | 47 | 23.05 |
| 28-Sep | 43 | 26.85 |
| 13-Oct | 38 | 31.45 |
| 28-Oct | 34 | 36.80 |
| 12-Nov | 30 | 42.70 |
| 27-Nov | 27 | 48.15 |
| 12-Dec | 26 | 51.85 |
| 27-Dec | 26 | 52.50 |
| 11-Jan | 27 | 49.80 |
| 26-Jan | 29 | 44.80 |
| 10-Feb | 33 | 38.90 |
| 25-Feb | 37 | 33.20 |
| 12-Mar | 42 | 28.25 |
| 27-Mar | 46 | 24.20 |
| 11-Apr | 50 | 21.00 |
| 26-Apr | 53 | 18.65 |
| 11-May | 55 | 17.00 |
| 26-May | 57 | 16.00 |
| 10-Jun | 58 | 15.50 |
| 25-Jun | 58 | 15.35 |
| 10-Jul | 58 | 15.60 |
| 25-Jul | 57 | 16.30 |
| 9-Aug | 55 | 17.60 |
| 24-Aug | 52 | 19.50 |

## Graphical representation

We need represent the results graphically by Figure (7) and Figure (8). Figure (7) represents the shadow length of the chosen building, and figure (8) is more general, because it represent the percent of shadow with the real height of any building.

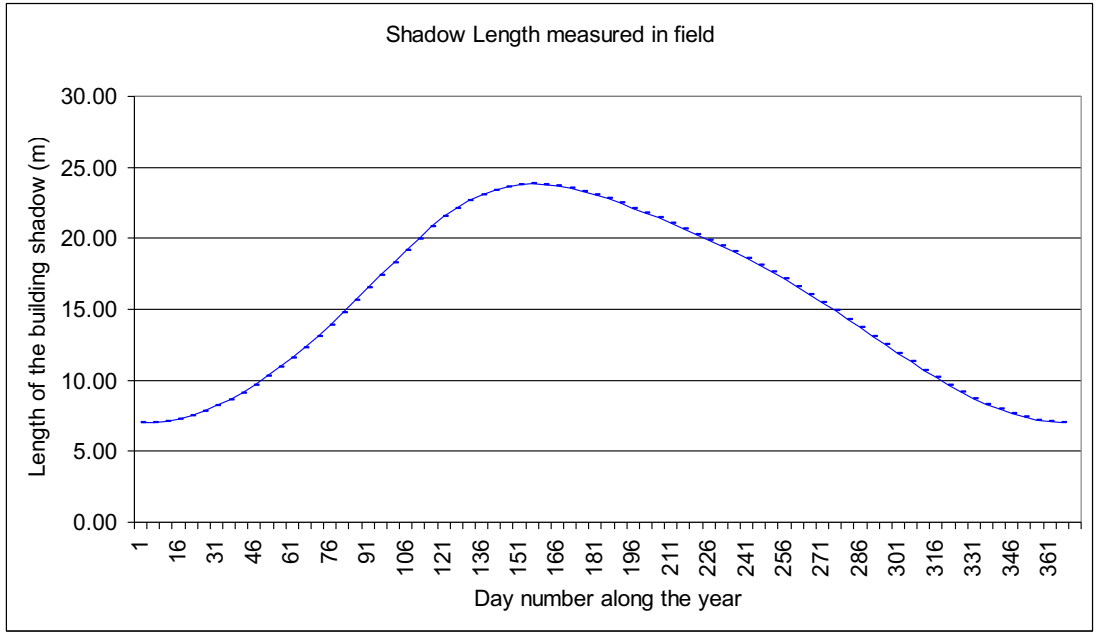

Figure (7) The change of shadow length during a year.

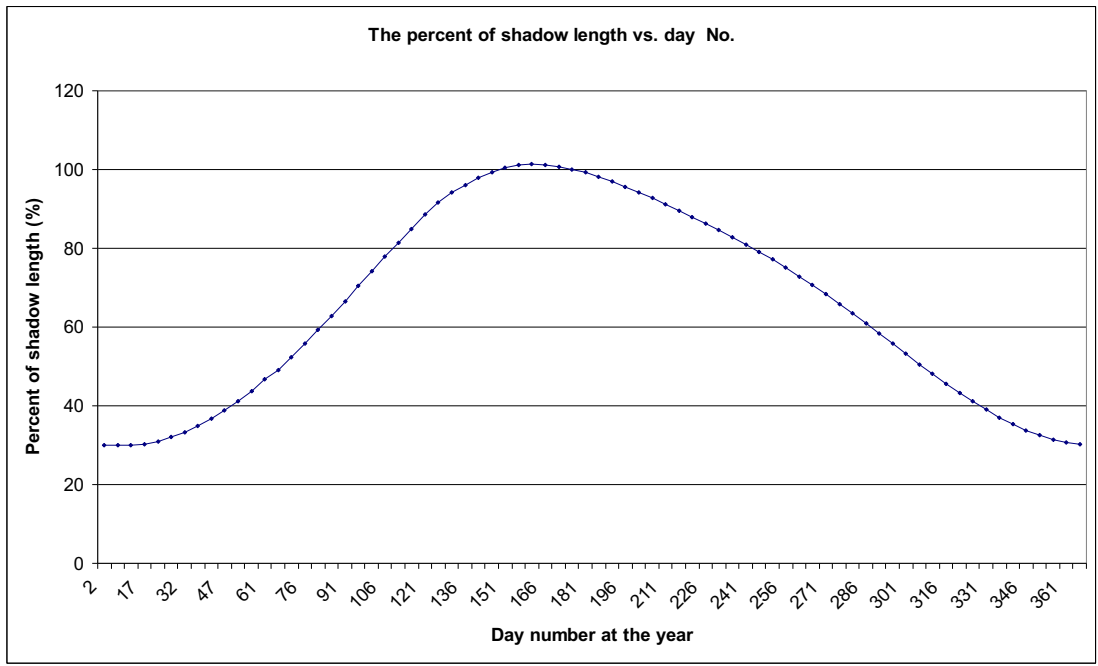

Figure (8) the percent of shadow length vs. day number along the year.

We depend the numbers that resulted from the astronomic equations for more accuracy possible.

The previous curve is similar to the normal distribution but there is a skew to the left, this need some modification and similar equation take into account this skew.

## The derived Gaussian equation

The Gaussian distribution is a common continuous distribution. Physical quantities and some independent processes often have distributions that are approximately normal [R. 14].

The normal Gaussian distributions are sometimes informally called the bell curves. We can add two Gaussian distribution to make superposition between the two equations and to describe some unsymmetrical tendency.

The general Gaussian form which is best fitting to the data distribution listed in tables (1 &), and figure (8) is

$$\text{SLP} = a_1 . e^{-\left(\frac{d_n - b_1}{c_1}\right)^2} + a_2 . e^{-\left(\frac{d_n - b_2}{c_2}\right)^2} \quad \dots \dots \dots \dots \dots (6)$$

Where:
SLP: is the Shadow length percent (table 2).

Coefficients (with 95% confidence bounds):
$a_1 =$ 99.22
$b_1 =$ 165
$c_1 =$ 130.7
$a_2 =$ 19.28
$b_2 =$ 329.6
$c_2 =$ 96.92

Goodness of fit:
SSE: 1961
R-square: 0.9913
Adjusted R-square: 0.9912
RMSE: 2.337

We used the best fitting tool in MATLAP program to find our equation with its constants. These constant can be varying with the change of imaging time of the satellite an the altitude of the regions.

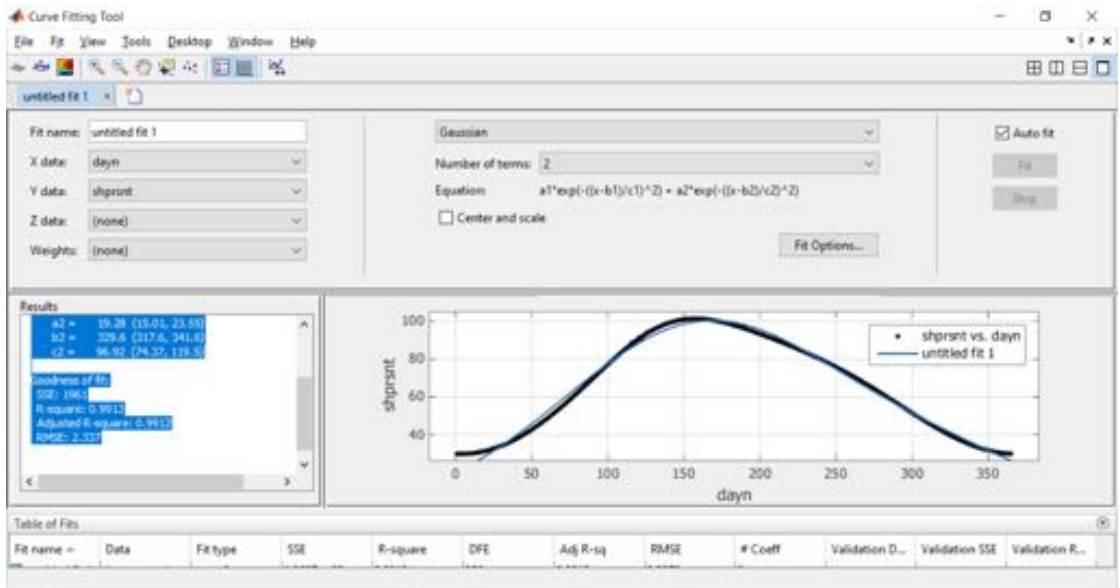
Figure (9) The best fitting tool in MATLAP program.

Now we check the resulted function, for example if we need to find the building height in 29/8/2017 ($d_n$=241) we apply the equation as below:

$$SLP = 99.22 * e^{-\left(\frac{241-165}{130.7}\right)^2} + 19.28 \cdot e^{-\left(\frac{241-329.6}{96.92}\right)^2}$$

$= 99.22 * 0.7131 + 19.28 * 0.433579$

$= 79.11\%$

The shadow length from table (1) is (20.31 m)

The real height so= 20.31/0.7911= 25.67 m

The real height of the building is 25.30 m

So, the error percent is= 1.4 %

By testing some other day numbers it is cleared the we can depend this accurate Gaussian equation to find the percent of shadow of any point in our city.

### 6. Discussion and conclusion
The derived empirical Gaussian equation in our work is useful equation to find the percentage of shadow for any building in the study region along the year, also it can be used to find the height of any building from the imaging date and its shadow length calculated from the satellite image.

We can also find the Gaussian equation for any other region to find the shadow percent of and known height building when it is needed to know the shadow extension on the street, architectural aims, or any other civil purposes.

The Gaussian equation provide practical calculation for any engineer or designer, and can be achieved by a simple hand calculator or a programmed Excel sheet instead of the astronomy equations.